\documentstyle[aps,twocolumn,epsf]{revtex}

\newcommand{\half}{\textstyle{\frac{1}{2}}}

\begin{document}
\title{Mixing property of triangular billiards}
\author{Giulio Casati$^{1}$, and Toma\v z Prosen$^{2}$}
\address{
$^{1}$International Center for the Study of Dynamical Systems,
Universita'degli Studi dell'Insubria,
via Lucini,3, I--22100 Como,\\
and Istituto Nazionale di Fisica della Materia and INFN, Unit\`a di
Milano, Italy,\\
$^{2}$Physics Department, Faculty of Mathematics and Physics,
University of Ljubljana, Jadranska 19, 1111 Ljubljana, Slovenia}
\date{\today}
\draft
\maketitle
\begin{abstract}
We present numerical evidence which strongly suggests that
irrational triangular billiards (all angles irrational with $\pi$) are
mixing. Since these systems are known to have zero Kolmogorov-Sinai
entropy, they may play an important role in
understanding the statistical relaxation process.
\end{abstract}

After the pioneering paper of Fermi, Pasta and Ulam\cite{FPU} and the
mathematical works of Kolmogorov
school, the modern ergodic theory can now account for the rich variety of
different statistical behaviours of classical dynamical systems. These
properties range from complete integrability to deterministic chaos.
The relatively few rigorous results so far available have provided a firm
guide for a large amount of analytical and numerical work which gave a
strong impulse to the field of nonlinear dynamics.
Needless to say, several problems remain to be solved. For example it is
known that mixing property guarantees correlations decay and relaxation to
statistical equilibrium. It is however not known whether or not this
property is sufficient for a meaningful statistical description of the
relaxation process or whether the strongest property of positive KS
entropy is required.
Even much less clear is the situation in quantum mechanics. For example,
to what extent the statistical distribution of energy levels of a quantum
conservative hamiltonian system, is related to the different properties in
the ergodic hierarchy is, as far as we know, an open question.

Examples of classically completely integrable or deterministic random
systems have been widely studied in the literature. However, to our
knowledge, there are no physical examples of systems which possess the
mixing property only (with zero KS entropy). In this respect, the best
candidates are billiards in 2-d triangles but, in spite of 30 years of
investigations, no definite statement can be made concerning their
dynamical properties\cite{katok}. Moreover, \cite{Gutkin} ``a
prevailing opinion in the mathematical community is that polygonal
billiards are never mixing, but this has not been established".

In this paper we provide strong numerical evidence that generic
triangular
billiards, with {\em all angles irrational} with $\pi$, are mixing.
This result can  play an important role in the
understanding of non-equilibrium statistical mechanics since
it fills a gap in the ergodic hierarchy. Moreover, since the
local dynamical instability in these systems is only linear,
they have zero algorithmic complexity. This means that, even though
it may be very difficult, their analytical solution is not
impossible in principle and this may prove to be very important
for the future development of nonlinear dynamics.

We consider the motion of a point particle, with unit velocity,
inside a generic triangular billiard (with all angles irrational
with $\pi$, in general). Let us introduce
cartesian coordinates $(x,y)$, and align one side of the
triangle along the $x-$axis ($y=0$), starting at the origin
$x=0$, and let us choose its length to be unity. We
denote the angle at the origin $(0,0)$ by $\alpha$ and the
angle at $(1,0)$ by $\beta$. Hence the other two sides of
the triangle are given by the eqs. $y=(\tan\alpha)x$, and
$y=(\tan\beta)(1-x)$.

We would like to draw the attention to the following fact: the
dynamics of a  point particle in a triangular billiard is
{\em equivalent} to the motion of three particles on a ring
with different masses,
$m_1$, $m_2$, and $m_3$.
Let the ring have circumference $1$.
The motion of 3-particles 1d gas with coordinates
$q_1,q_2,q_3$ ($q_1 \le q_2 \le q_3 \le q_1 + 1$) is governed
by the hamiltonian $H=\half m_1 \dot{q}_1^2 + \half m_2
\dot{q}_2^2 + \half m_3 \dot{q}_3^2$, with elastic collisions
at $q_1 = q_2$, $q_2 = q_3$, and $q_3 = q_1 + 1$.
Introducing the notation
$$\vec{r} = (\sqrt{m_1}q_1,\sqrt{m_2}q_2,\sqrt{m_3}q_3)$$
and  the orthogonal transformation
\begin{eqnarray*}
x &=& \frac{1}{\sqrt{(m_1+m_2)M}}
(-\sqrt{m_1 m_3},-\sqrt{m_2 m_3},m_1 + m_2)\cdot\vec{r},\\
y &=& \frac{1}{\sqrt{m_1 + m_2}}
(-\sqrt{m_2},\sqrt{m_1},0)\cdot\vec{r},\\
z &=& \frac{1}{\sqrt{M}}
(\sqrt{m_1},\sqrt{m_2},\sqrt{m_3})\cdot\vec{r},
\end{eqnarray*}
with $M=m_1+m_2+m_3$, the hamiltonian  writes:
$H=\half(\dot{x}^2+\dot{y}^2+\dot{z}^2)$
 where, as can be checked by
straightforward calculations, the motion in the $(x,y)-$plane
 is bounded
by  specular reflections from the sides of the triangle
 with  angles
\begin{equation}
\tan\alpha=\sqrt{\frac{m_2 M}{m_1 m_3}},\;
\tan\beta =\sqrt{\frac{m_1 M}{m_3 m_2}},\;
\tan\gamma=\sqrt{\frac{m_3 M}{m_2 m_1}}.
\label{eq:angles}
\end{equation}
The motion in the $z-$coordinate trivially separates and
corresponds to the center-of-mass motion in the 3-particles 1d gas.

Notice, however, that  triangular billiards with  one
angle  larger than $\pi/2$ cannot be related to
the 3-particles 1d gas, since from the above formula it
follows that $\alpha,\beta,\gamma < \pi/2$.
A particular case is given by the isosceles triangle which is
dynamically equivalent to the  right triangle billiard, with e.g.
$\gamma=\pi/2$. The
latter corresponds to $m_3=\infty$, i.e. to the 1d motion
of two particles with masses $m_1$, $m_2$ between
hard walls\cite{Kornfeld}.

The dynamics of triangular billiards can be
divided into three classes:\\
(A) {\em All} angles rational with $\pi$. The dynamics
of such triangles is not ergodic; in fact it is
{\em pseudo-integrable}, i.e. it possesses 2d invariant surfaces
of high genus in 4d phase space.\\
(B) {\em Only one} angle  rational with $\pi$. Such are
generic right triangles, for example.
In recent numerical experiments evidence has been given that
right, irrational, triangular
billiards are ergodic and weakly mixing \cite{Artuso}.\\
(C) {\em All} angles {\em irrational} with
$\pi$.
Surprisingly, to the best of our knowledge, this generic
class of triangles has been somehow overlooked in previous
numerical studies and will be the main
object of the present paper.
It is within this class that one may now hopefully look for
ergodic and mixing behaviour.

From the rigorous point of view not
much is known beyond the fact that the
set of ergodic triangles is a dense $G_\delta$
(intersection of a countable number
of dense open sets) in a suitable topology.
We recall that a dynamical system
$T^t : \vec{x}(0) \rightarrow \vec{x}(t)$ with invariant
measure $\mu(\vec{x})$ in phase space ${\cal M}$ is mixing
if, for any $L^2$ pair of observables in phase space, their time
correlation function asymptotically vanishes
$$\lim_{t\rightarrow\infty}
\left[\int\limits_{\cal M} d\mu f(T^t\vec{x}) g(\vec{x})
- \int\limits_{\cal M} d\mu f(\vec{x})
  \int\limits_{\cal M} d\mu g(\vec{x})
\right]=0.
$$
The map $T^t$ may represent a continuous flow ($t$ real) or
a discrete map ($t$ integer). In this paper we shall mainly
consider the dynamics given by a discrete Poincar\' e map which
corresponds to the collisions of the orbits with the
horizontal side $y=0$. The reduced phase space --- surface
of section (SOS) --- is a rectangle,
parametrized by the coordinate $0 \le x \le 1$
and by the corresponding canonical momentum
$-1 \le p_x=\sin\vartheta \le 1$
($\vartheta=$ angle of incidence),
with the invariant measure $d\mu = dx dp_x$.

As a first step we have  performed an accurate
check of ergodicity. We have done this in two independent
ways:

(i) By dividing the phase space (SOS) in a large number
$N=N_1\times N_1$ of cells and then computing the number
$n(t)$ of cells which are visited by a single orbit up to
discrete time t \cite{Robnik}.
Then we computed the average relative measure
$r(t) = \langle n(t)/N \rangle$ of visited
phase space up to time $t$
where $\langle.\rangle$ denotes phase space
average over sufficiently many randomly chosen initial conditions.
As it is known, for the so-called {\em random model} of completely uniform
(ergodic) and random dynamics, one can derive the simple scaling
law \cite{Robnik}
\begin{equation}
 r(t)= r_{RM}(t) = 1 - \exp(-t/N).
\label{eq:RM}
\end{equation}
Expression  (\ref{eq:RM}) (for arbitrary but sufficiently
fine mesh $N$) should be considered as a sufficient but not
necessary condition for ergodicity.
In fact, the dynamics would obey the law (\ref{eq:RM})
only when the system does not possess any
non-trivial time scale.
In Fig.1 we show the results of numerical computations
for triangles of class (B) and (C).
We take for example a right triangle B with angles
$\alpha=(\sqrt{5}-1)\pi/4,\beta=\pi/2-\alpha,\gamma=\pi/2$
and a generic triangle C with angles
$\alpha=(\sqrt{2}-1)\pi/2, \beta=(\sqrt{5}-1)\pi/4,
\gamma=\pi-\alpha-\beta.$
It is seen that there is a drastic difference between the two
cases: the generic triangle C excellently follows the law (\ref{eq:RM}),
whereas the right triangle B, even though ergodic,\cite{Artuso} strongly
deviates
and
explores the phase space extremely slowly, with
a hierarchy of long time-scales related to a strong sticking
of the orbit in momentum ($p_x$) space.

\begin{figure}[htbp]
\hbox{\hspace{-0.1in}\vbox{
\hbox{
\leavevmode
\epsfxsize=3.6in
\epsfbox{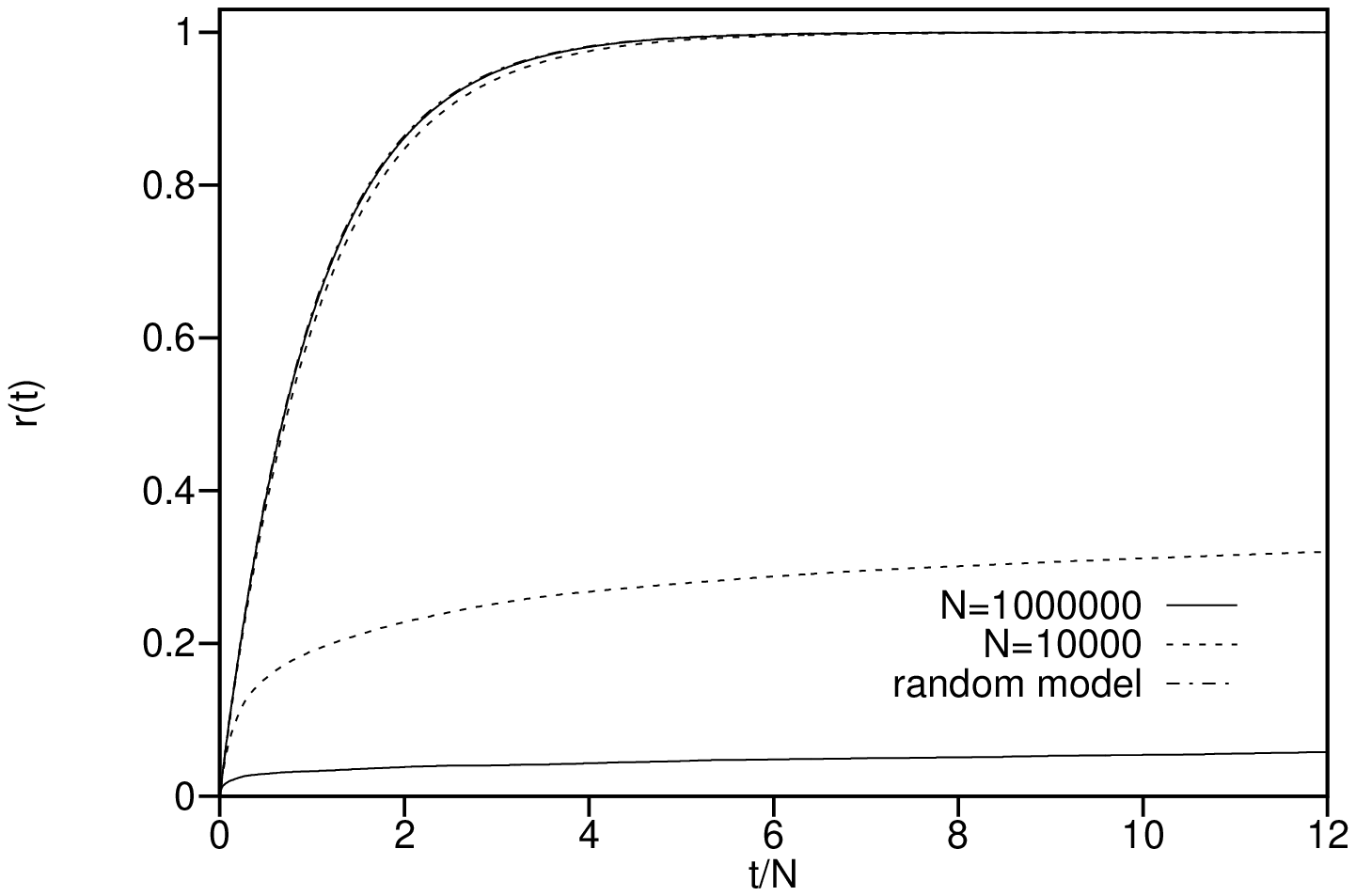}}
\vspace{-2.13in}
\hbox{\hspace{0.9in}
\vbox{
\leavevmode
\epsfxsize=2.45in
\epsfbox{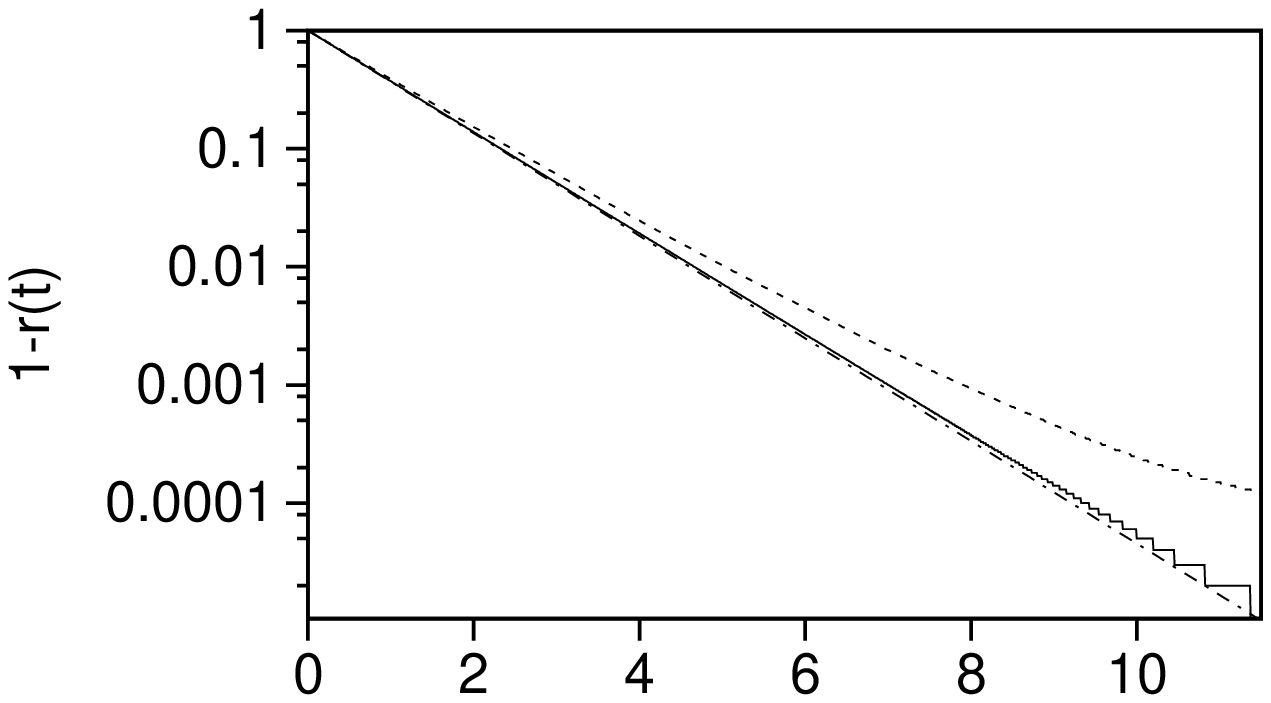}}
\vspace{0.92in}
}}}
\caption{
The relative measure $r(t)$ of the visited phase space as a function
of discrete time $t$. The full and dashed curves
refer to different discretizations of phase space, namely to
$N=10^6$ and $N=10^4$ cells, respectively.The two upper
curves refer to the generic
triangle C whereas the two lower curves refer to the golden right
triangle B (see text for details).
The random model, expr.(2), is given by the dotted-dashed
curve which is almost indistinguishable from the numerical
curve for the triangle C at $N=10^6$.
In order to better display the excellent agreement with
expr.(2), we plot in the inset the function $1-r(t)$ for the
generic triangle C in the
semi log-scale. At $N=10^6$, and $N=10^4$, averages
over $200$, and
$4000$ orbits with randomized initial conditions have been used,
respectively.
}
\label{fig:1}
\end{figure}

(ii) By comparing the {\em time averaged} correlation
function for a single, but very long orbit,
$ 
C^t_a(t)=\lim_{T\rightarrow\infty}\frac{1}{T}\sum_{n=1}^T a(n)a(n+t)
$
with the phase averaged
correlation function 
$C^p_a(t)=\langle a(0)a(t)\rangle$ which is computed by monte-carlo
averaging over many short orbits with different randomized
initial conditions. In Fig.2 we plot the difference between the two curves
for the generic triangle C which turns out to be of the same order as the 
statistical error. 

\begin{figure}[htbp]
\hbox{\hspace{-0.1in}\vbox{
\hbox{
\leavevmode
\epsfxsize=3.6in
\epsfbox{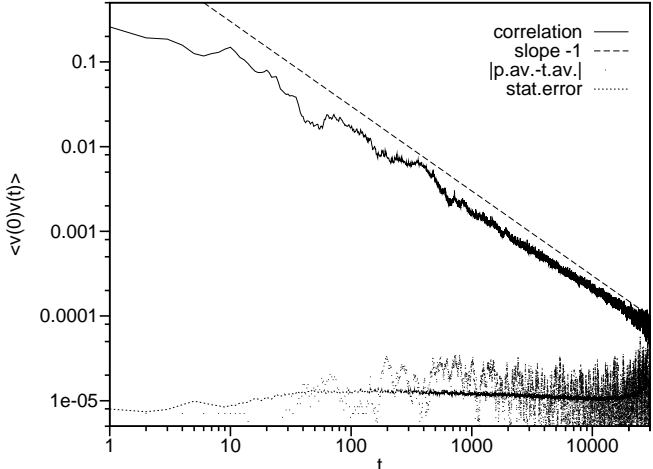}}
}}
\caption{
The velocity autocorrelation function
$C^p_{p_x}(t)$ in log-log scale for the generic
triangle C (full curve), which is obtained as the average
over $1.2\cdot 10^6$ different orbits of length $32768$.
The dashed line gives the
slope $-1.00$.
The dotted curve gives the estimated statistical error of the
correlation function, while the dots mark the difference
between the phase averaged and the time averaged correlation
function $|C^p_{p_x}(t)-C^t_{p_x}(t)|$ where the time average
is computed from a single orbit of length
$ T=  32768\cdot 1.2\cdot 10^6$
with initial condition $x_0=0.23456,p_{x0}=0.34567$. The 
statistical error is estimated as the standard deviation
of a sequence of $ M=1000$ partial averages of length T/M each.}
\label{fig:2}
\end{figure}

\noindent
The above numerical results are surprisingly much
more clear than expected; they demonstrate ergodic
behaviour and they make reasonable to  expect
mixing behaviour also. We now turn our attention to the latter
question.

We have performed extensive numerical computations
of autocorrelation functions of  different observables,
namely, momentum (horizontal component of velocity) $v=p_x$,
symmetrized position $x'=2x-1$, and also characteristic
functions of sets in phase space.
In all cases we have found clear numerical evidence of
{\em power-law decay} of correlation functions over about 4
orders of magnitude. For a given triangle, correlation functions
of different observables, decay with the same
empirical exponent which is typically very close to $-1$.
In Fig.2 we show the velocity correlation function of the
triangle C, which decays with exponent with the fitted value 
$-\sigma=-0.90\pm 0.02$. However, since the triangle C is not 
too far from the right triangle we have chosen another, 
similar generic triangle $D$ with the irrational angles 
$\alpha = (\sqrt{2}-1)\pi/2, \beta=1, \gamma=\pi-\alpha-\beta$ and 
show in Fig.3 its velocity and position correlation function 
which decay with empirical exponent $-\sigma=-0.94\pm 0.04$.
We have performed extensive numerical calculations 
of correlation functions for many different generic triangles,
and the exponent of decay have always been very close to $-1$.
There are even cases where the fitted decay exponent seems to be 
slightly below $-1$ ($\sigma > 1$), 
however in such cases, correlation functions
appear much more noisy and it is difficult to make precise statements.

\begin{figure}[htbp]
\hbox{\hspace{-0.1in}\vbox{
\hbox{
\leavevmode
\epsfxsize=3.6in
\epsfbox{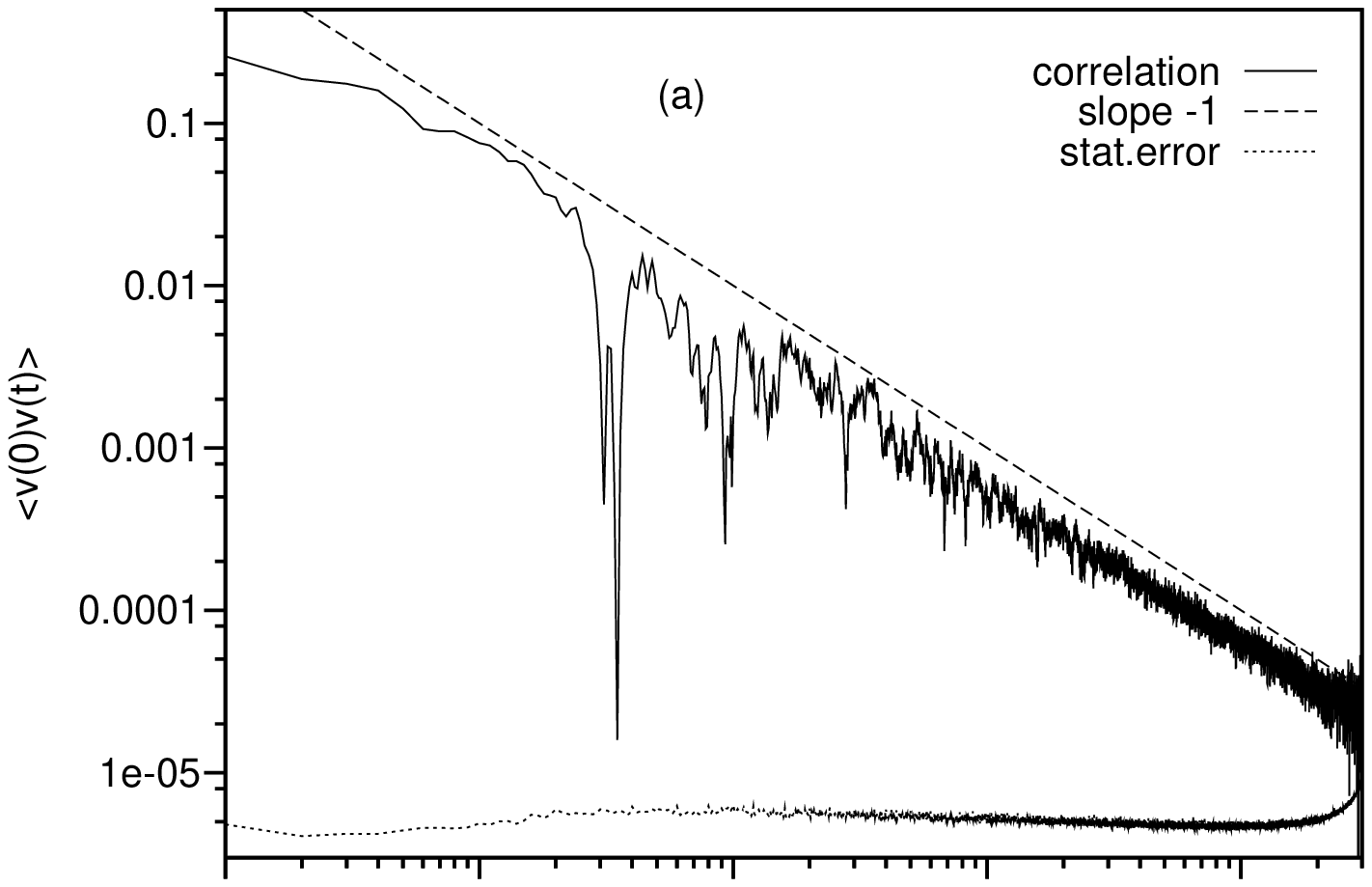}}
\vspace{-0.12in}
\hbox{
\leavevmode
\epsfxsize=3.6in
\epsfbox{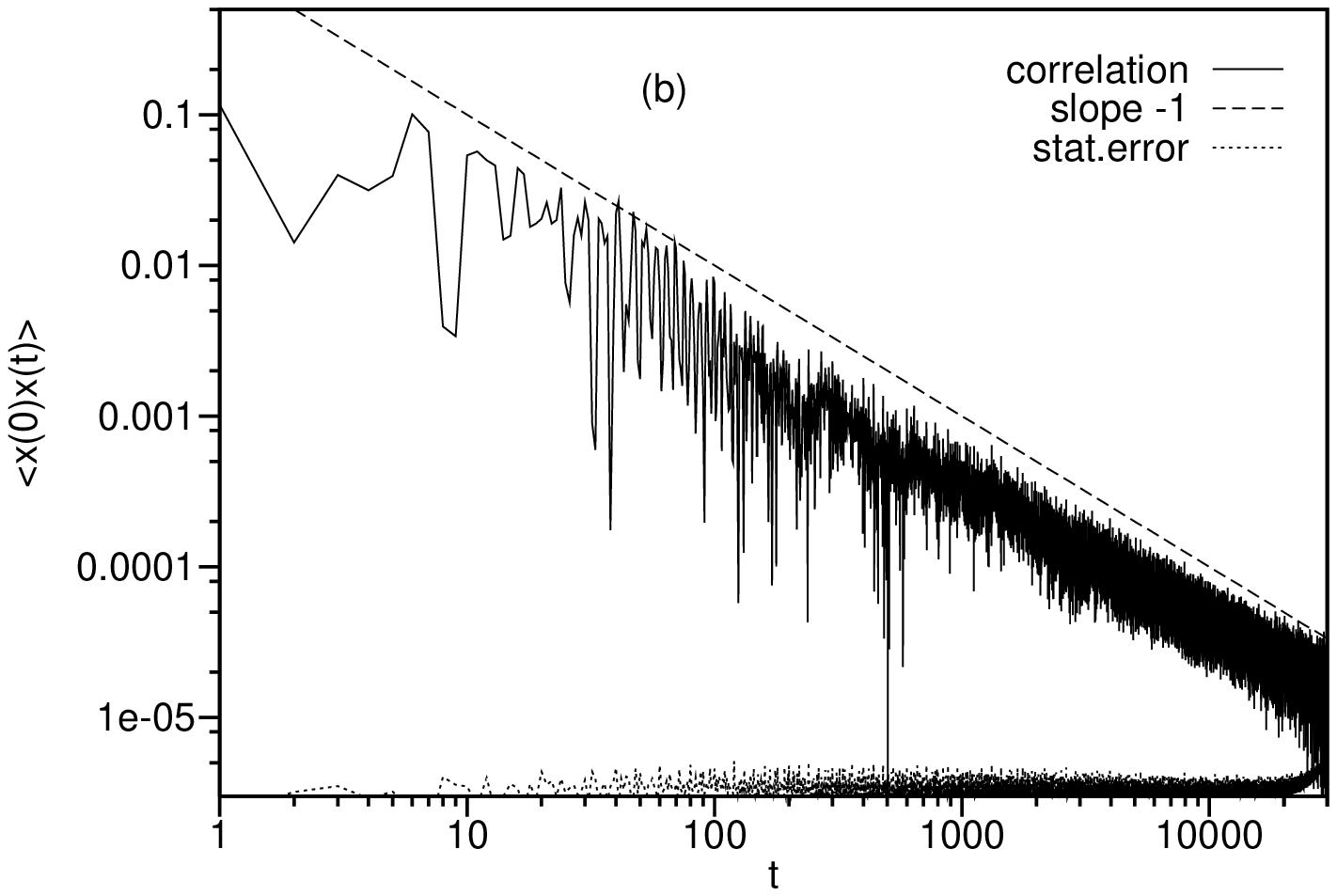}}
}}
\caption{
The velocity (a) and position (b) correlation function,
$C^t_{p}(t)$ and $C^t_{x'}(t)$, for an orbit of length $10^{11}$
in the generic triangular billiard D. Initial condition:
$x_0 = 0.23456, p_{x0} = 0.34567$.
}
\label{fig:3}
\end{figure}

Finally, we have performed a different and powerful statistical
test, namely we have computed the Poincare' recurrences or the
return time statistics, i.e. the probability $P(t)$ for an orbit not
to stay outside a given subset ${\cal A}\subset {\cal M}$ for a time
longer than $t$. It has been conjectured \cite{Dima,Karney}
(see also\cite{Artuso2}
for accurate definition and further references) that the integrated
probability $P_i(t) = \sum_{t'=t}^\infty P(t')$ should be intimately
connected to the correlation decay. More precisely, if the correlation
function decays as a power-law with exponent $-\sigma$, then the
integrated return probability should decay with a similar exponent
$P(t) \sim t^{-\mu-1}, P_i(t) \sim t^{-\mu}$,  where $\mu =\sigma$.
In Fig.4 we show the recurrence probability $P(t)$ and the
integrated probability $P_i(t)$ computed w.r.t. half-space
${\cal A}=\{(x,p_x);p_x>0\}$ for the same data as in Figs.2,3
(triangular billiard C). We have found that numerical results are
perfectly consistent with $\mu=\sigma=1$, i.e. with $1/t^2$ decay
of return probability $P(t)$.

To summarize, the numerical obtained value of exponent $\sigma$ is close
to -1 but in some cases is slightly less and in other cases slightly
larger. We attribute this fact to long transient times which are caused by
the intricate interplay of arithmetic properties of angles $\alpha$,
$\beta$, $\gamma$ and by the presence of periodic orbits. In
this connection we would like to mention that we have numerically
investigated the existence of periodic orbits and we have found that their
number increases very slowly with their period. The number of
non-equivalent periodic orbits with lengths up to $l$ (collisions with
the boundary) typically increases {\em slower} than $l$.

\begin{figure}[htbp]
\hbox{\hspace{-0.1in}\vbox{
\hbox{
\leavevmode
\epsfxsize=3.6in
\epsfbox{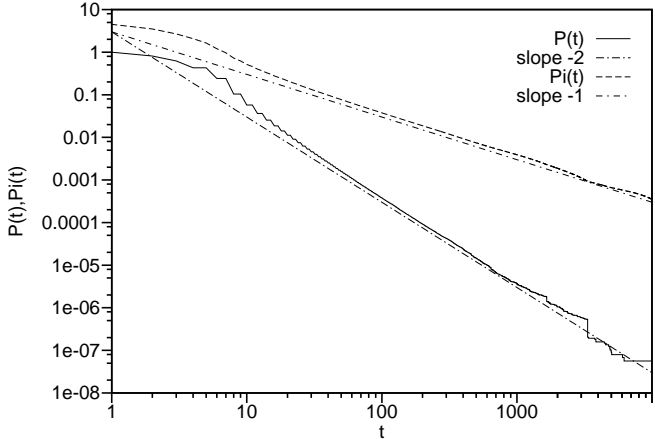}}
}}
\caption{
The return probability $P(t)$ (full curve), and the integrated
return probability $P_i(t)$ (dashed curve), for the same orbit
(triangle D) as in Fig.3. The dashed-dotted lines have slopes $-2$ and
$-1$ respectively.}
\label{fig:4}
\end{figure}

A qualitative argument which leads to the $1/t$ decay of correlations reads
as follows: let us consider a fixed, sufficiently small, region 
${\cal A}$ in
phase space around a periodic orbit and study the $l$ iterates
of the Poincar\' e map, where $l$ is the period of the orbit. 
Due to the linear (in)stability, the
orbit which starts in a small square $a_\epsilon$ of side $\epsilon$
centered on the periodic orbit and contained in ${\cal A}$, will
remain in ${\cal A}$
for a time $t \ge 1/\epsilon$. It follows that the probability $P(1/\epsilon)$
for an orbit to stay in the region ${\cal A}$ for a time $t \ge 1/\epsilon$ is
proportional to the probability of the orbit to visit the
square $a_\epsilon$. Now, the linear instability implies that an orbit
which enters the square $a_\epsilon$ will remain inside $a_\epsilon$ for a
constant time which does not depend on $\epsilon$ as 
$\epsilon\rightarrow 0$. As a consequence, 
from the Liouville theorem it follows that the probability for an orbit
to visit the square $a_\epsilon$ will be proportional to its area
$\epsilon^2$. This is the probability $P(1/\epsilon)$ to remain in the
region ${\cal A}$ for a time $t \ge 1/\epsilon$. This leads to the relation
$P(t) \propto 1/t^2$ as confirmed by numerical results.

Since the question discussed in this paper is a very delicate one,
we have put particular attention to the accuracy of our numerical
computations. Among the  several different tests, we have
 developed two independent
computer codes, one based on  the billiards dynamics in a 2d
plane, and the other based on the three particles 1d gas dynamics:
the results agree.
Moreover, since the instability here is only linear in time, and
the machine precision is $\sim 10^{-16}$, the numerical errors are
always below the statistical errors.

The central question is to what extent the results presented
here can be considered as a definite or convincing evidence
for the mixing property of generic triangular billiards.
The correlation decay of some particular class of functions
is certainly compatible with a weaker property than mixing.
However, in this paper we have compared the correlations decay
of different variables (velocity, position, characteristic functions of
phase space sets etc). We have also considered
the dynamics given by the discrete Poincare map
relative to different sides of the same triangle: all these correlations
exhibit a power law decay with exponent very close to $-1$, and make
very plausible the conjecture that correlation functions in generic 
triangles decay as $\propto 1/t$.
Of particular significance is the behaviour of the return probability
$P(t)$ which decays with the expected power law $\propto 1/t^2$. 
All the
above results (including those described in Fig.1) lead to the
conclusion that, outside any reasonable doubt, generic irrational triangles
are mixing.
The particular case of isosceles or right triangles is much less clear.
The behaviour of correlations in such case is noisy and the exponent of the
decay is too small to allow for any definite conclusion.

The fact that generic triangles have zero Kolmogorov-Sinai entropy
and yet they have a very nice mixing behaviour without any time scale,
may prove to be very useful for understanding the dynamical basis
of the relaxation process to the statistical equilibrium.
Moreover, the analysis of their quantum behaviour will contribute to
the current efforts for the construction of a statistical theory of quantum
dynamical systems.

The authors acknowledge the hospitality of Centre Internacional de
Ciencias, a.c, Cuernavaca, Mexico, where the major part of this work has
been performed, and fruitful discussions with F. Leyvraz.
T.P. gratefully acknowledges the financial support by
the ministry of science and technology of R Slovenia.


\begin{thebibliography}{99}
\bibitem{FPU} E. Fermi, J. Pasta and S. Ulam, Los Alamos
preprint LA-1940, (1955).

\bibitem{katok}E. Gutkin and A. Katok, Lectures Notes in Math. vol
1345, p.163-176
(Springer,Berlin, 1989).

\bibitem{Gutkin}E. Gutkin, J. Stat. Phys. {\bf 83},7 (1996).

\bibitem{Artuso} R. Artuso, G. Casati, and I. Guarneri,
Phys. Rev. E {\bf 55}, 6384 (1997).

\bibitem{Kornfeld} P. Kornfeld, S. V. Fomin, Y. G. Sinai,
{\em Ergodic Theory} (Springer, Berlin, 1982).

\bibitem{Dima}B.V.Chirikov and D.L. Shepelyansky,
 INP 81-69,1981 (unpublished) (in Russian)(English
 translation: 
 Report No PPPL-TRANS-133, Princeton, 1981.

\bibitem{Karney} C.F.F. Karney, Physica D{\bf 8}, 360 (1983).

\bibitem{Artuso2} R. Artuso, {\em Correlation decay and
return time statistics}, Physica D (to appear).

\bibitem{Robnik} M. Robnik, J. Dobnikar, A. Rapisarda,
T. Prosen, and M. Petkov\v sek, J. Phys. A {\bf 30}, L803
(1997).

\end{thebibliography}
\end{document}